\documentstyle[12pt,psfig]{article}
\def\reference{\parskip 0pt\par\noindent\hangindent 0.5 truecm}

\begin{document}
\title{A very rapid Extreme Scattering Event in the IDV source
0954+658}
\author{Giuseppe Cim\`o,
\and
 T. Beckert, 
 T. P. Krichbaum, 
 L. Fuhrmann, 
 A. Kraus,\\
 A. Witzel, 
J. A. Zensus
} 
\date{}
\maketitle
{\center
Max-Planck-Institut f\"ur Radioastronomie, Auf dem H\"ugel 69,
Bonn, Germany\\gcimo@mpifr-bonn.mpg.de\\[3mm]
}
\begin{abstract}
\noindent
Extreme Scattering Events (ESEs) are dramatic variations of
the flux density at
Gigahertz frequencies caused by ray path distortions within an isolated
inhomogeneity (``pla\-sma lens'') in the interstellar medium.
These events are characterized by a deep flux density 
minimum in the light curve with, in some cases,
surrounding maxima. The variability time scales range from weeks
to months. These phenomena show a strong frequency dependence, 
in which the variability amplitudes increase with wavelength.\\
During an Intraday Variability (IDV) monitoring project (March 2000), a feature
resembling an ESE-like event appeared in the variable light curve
of 0954+658, however with a time scale of less than two days. We
will discuss this effect and its
implications for a better description of the interstellar medium.
\end{abstract}

{\bf Keywords:}
Quasars: Individual (0954+658)
                -- ISM: Structure, Scintillation
\bigskip
%
%
\section{Introduction}
The flux densities of Intraday Variable radio sources (Heeschen et al. 1987)
show significant variations within time scales of a few hours to $\sim$2 days. 
Due to small source size, either refractive interstellar scintillation (RISS) 
or source intrinsic effects or a mixture of both are possible causes 
for Intraday Variability (IDV, see Wagner \& Witzel 1995). 
In addition, peculiar flux density variations with a deep minimum symmetrically
surrounded by enhanced flux appearing on time scales of weeks to months 
were observed in some sources and Fiedler et al. (1987) explained
such variations as due to strong scattering by isolated structures in 
the interstellar medium. 
In a statistical description of this phenomenon, Fiedler et
al. (1994) suggested that the identification of an ESE is difficult if
the amplitude and the time scale of the variations are comparable to
other possible origins of variability and RISS may be confused with
small amplitude and very rapid ESE.\\
In the following we will show that we have probably detected an
ESE-like event on short time scales in the multifrequency light 
curve of the intraday
variable BL~Lac object 0954+658 
with the 100\,m radio telescope at Effelsberg in March 2000.\\
0954+658 was the first source in which a variability pattern
resembling an ESE was seen (Fiedler et al. 1987), indicating the
presence of a strongly turbulent interstellar medium along the line of sight.
However, the ESE of 1981 shows variations on a time scale of
$\sim18$ weeks, much longer than the effect seen in
fig.\ref{figlabel0}.

\section{Observations and Data Analysis}

\begin{figure}[h!]
\begin{center}
\vspace{-1cm}
\hspace{2cm}
\psfig{file=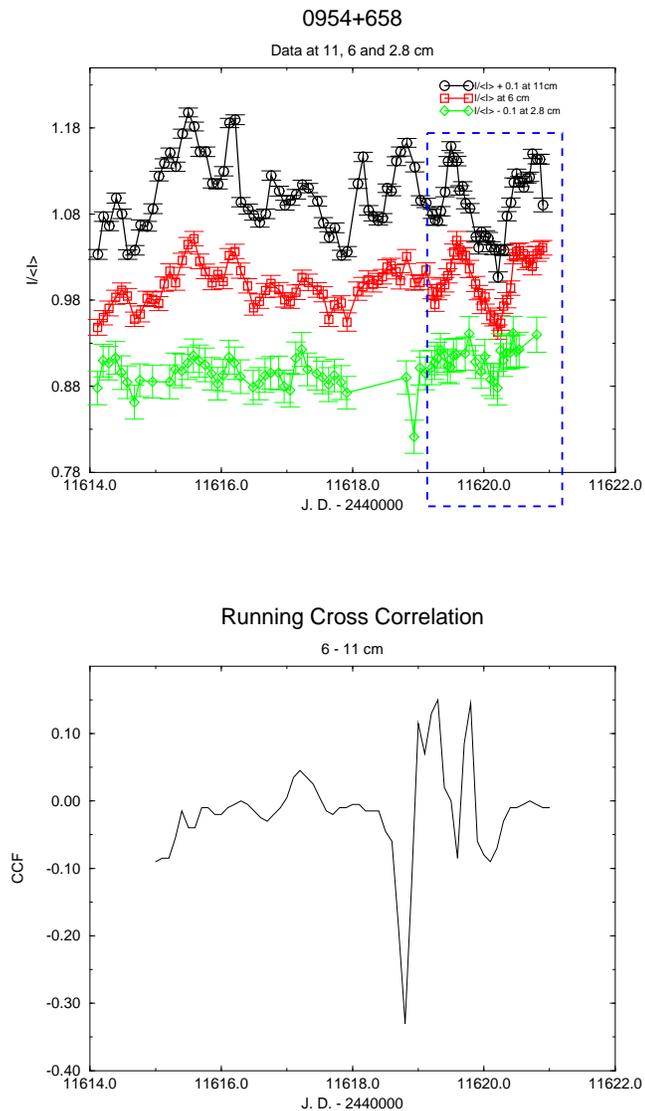,height=16cm,angle=270}
\caption{\textit{Top:} Light curve of 0954+658 at 11, 6 and
2.8\,cm. The dashed box highlights the ESE-like
event. \textit{Bottom:} Running Cross Correlation Function (CCF),
where we plot time lags versus time. We were not able to perform a
proper CCF analysis for the last hours of the event because it happened 
at the end of our observing period.}
\label{figlabel0}
\end{center}
\end{figure}

\begin{figure}[h!]
\begin{minipage}[t]{7cm}{
\psfig{file=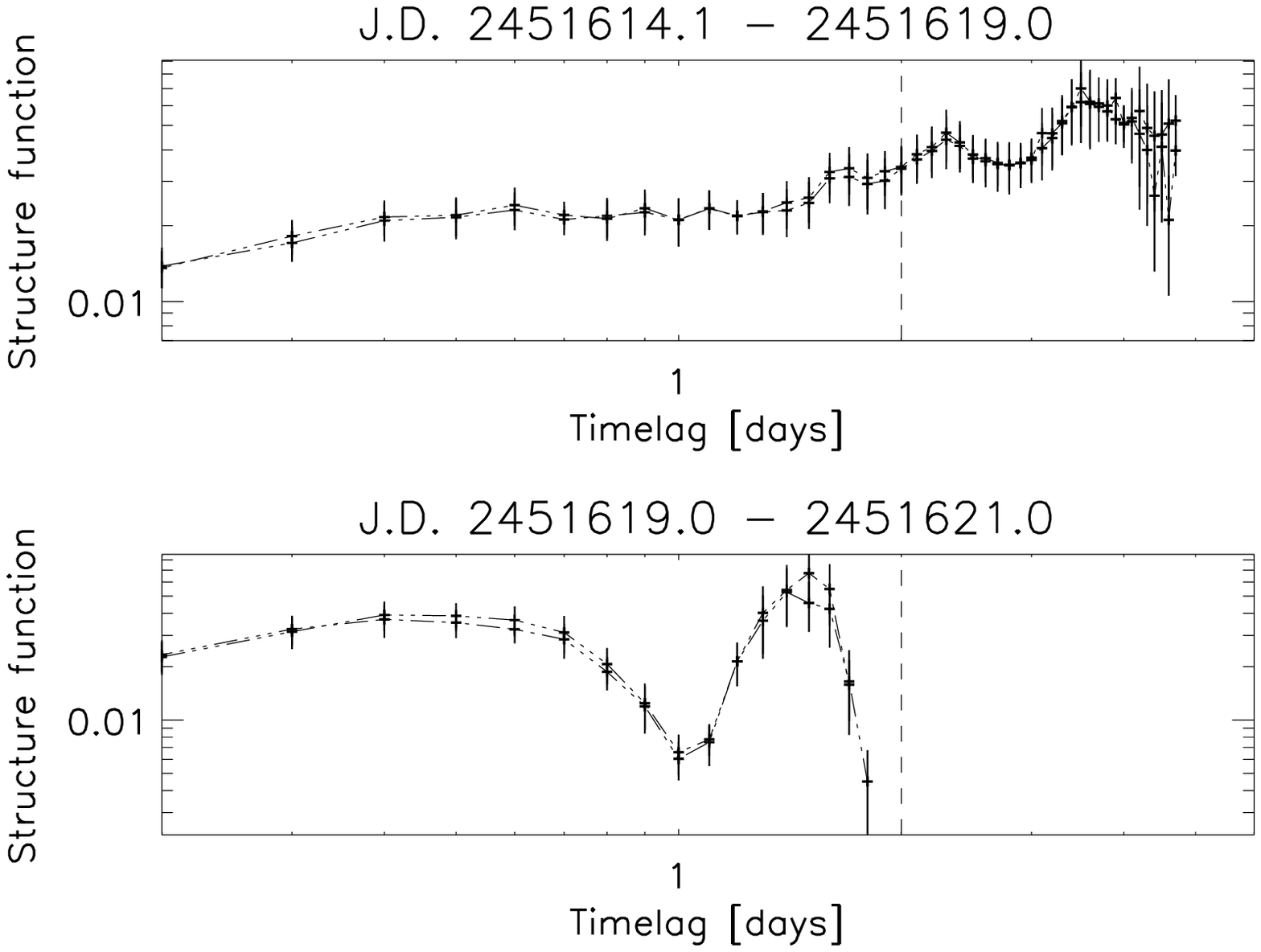,height=6cm}
\vspace{-0.7cm}
}
\end{minipage}
~~~~
\begin{minipage}[t]{7cm}{
\psfig{file=acf3.ps,height=5.5cm,angle=270}
\vspace{-0.7cm}
}
\end{minipage}
\caption{Structure functions and autocorrelation functions for the
polarized flux density at 6\,cm (normalized values). The minimum at 1\,day
(structure function at the bottom) is preserved when a longer time
interval is used. This minimum directly reflects the secondary maxima in the
autocorrelation function (dashed line).}
\label{figlabel2}
\end{figure}

In March 2000, we performed a radio-optical campaign to study possible 
correlations of IDV sources between different bands of the
electromagnetic spectrum.
The data, shown in this paper, were taken at 11, 6
and 2.8\,cm during part of this campaign (from March $10^{th}$ to
$17^{th}$) using the 100\,m radio telescope of the
Max-Planck-Institut f\"ur Radioastronomie (MPIfR) in Effelsberg. The data 
reduction was performed using CONT2, a task of the standard software package
TOOLBOX of the MPIfR. Observations of non-variable sources assured a
reliable flux density calibration (accuracy $\sim0.5\%$) allowing us to
correct for instrumental and atmospheric effects.
(Details of this data analysis can be found in Quirrenbach et al. 1992).\\
In the light curve (fig.\ref{figlabel0}, upper part) of the BL~Lac object
0954+658 ($m_V=17$ mag, z=0.367; Padovani \& Giommi 1995, Stickel et
al. 1993), we found a systematic change of the time lag between
adjacent frequencies. 
The first part of the light curve does not show any time lag between
the three frequencies while in the last part we observe rapid and (in
time) delayed changes: 
at J.D.$\sim2451619.6$, we observed a systematic time lag in the sense
that the flux density at longer wavelengths peaks before the flux
density at shorter wavelengths ($t_{11\,\mbox{cm}} < t_{6\,\mbox{cm}}
< t_{2.8\,\mbox{cm}}$). At J.D.$\sim2451620.8$, the situation reversed 
and the shorter wavelength was peaking before the longer wavelength.\\
To quantify these systematic variations of the time lag, we
performed a running cross correlation, 
i.e. time lags between 11 and 6\,cm 
versus time (see fig.\ref{figlabel0}, \textit{bottom}). The first part
of the plot shows a flat (centered in zero) pattern indicating no 
time lag between 6 and 11\,cm. During the
last 2 days of observations, a 
different behaviour is seen. 
We applied our model (see below) to the part of the light curve
where the time lag starts to be significantly different from zero.
Moreover, the cross correlation functions between 11 and 6\,cm
for the first and the last two days of observation show 
quite different behaviour. This again indicates that after
J.D. 2451619.6, the variability pattern and the time lag between
frequencies changed.\\
Analysis of the polarization gives further 
evidence for a different variability pattern before
J.D.$\sim$2451619.6: we noted that the polarized flux density 
variations after J.D.$\sim2451619.6$
are more pronounced and faster than before. Using a structure function 
analysis, we are able to quantify this change in the polarization
characteristics of the source. The results are showed in 
fig.\ref{figlabel2}, where we plot structure functions and 
autocorrelation functions for the 2  different time intervals at 6\,cm. 
The typical time scale of the variations changes from $\ge1.5$ to
$\sim$0.4\,days.

\begin{figure}[t]
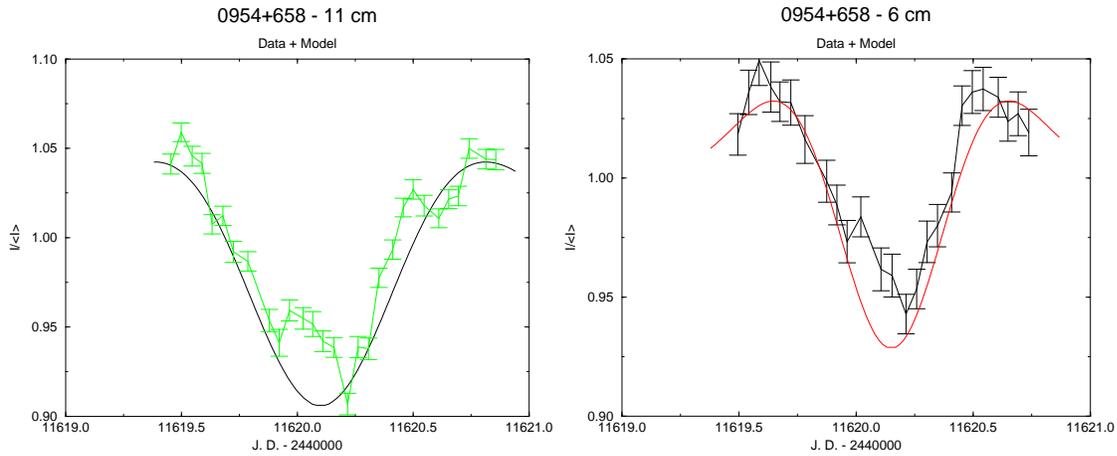

\begin{minipage}[t]{7cm}{
\psfig{file=11cmmodel+data.ps,height=6cm,angle=270}
\vspace{-0.7cm}
}
\end{minipage}
~~~~~~~~~~~~~~~~~~~
\begin{minipage}[t]{7cm}{
\psfig{file=6cmmodel+data.ps,height=6cm,angle=270}
\vspace{-0.7cm}
}
\end{minipage}
\vspace{-0.7cm}
\caption{Normalized flux density model light curves and normalized
data at 11 and 6\,cm}
\label{figlabel1}
\end{figure}

\begin{figure}[h!]
\begin{minipage}[t]{7cm}{
\hspace{-0.6cm}
\psfig{file=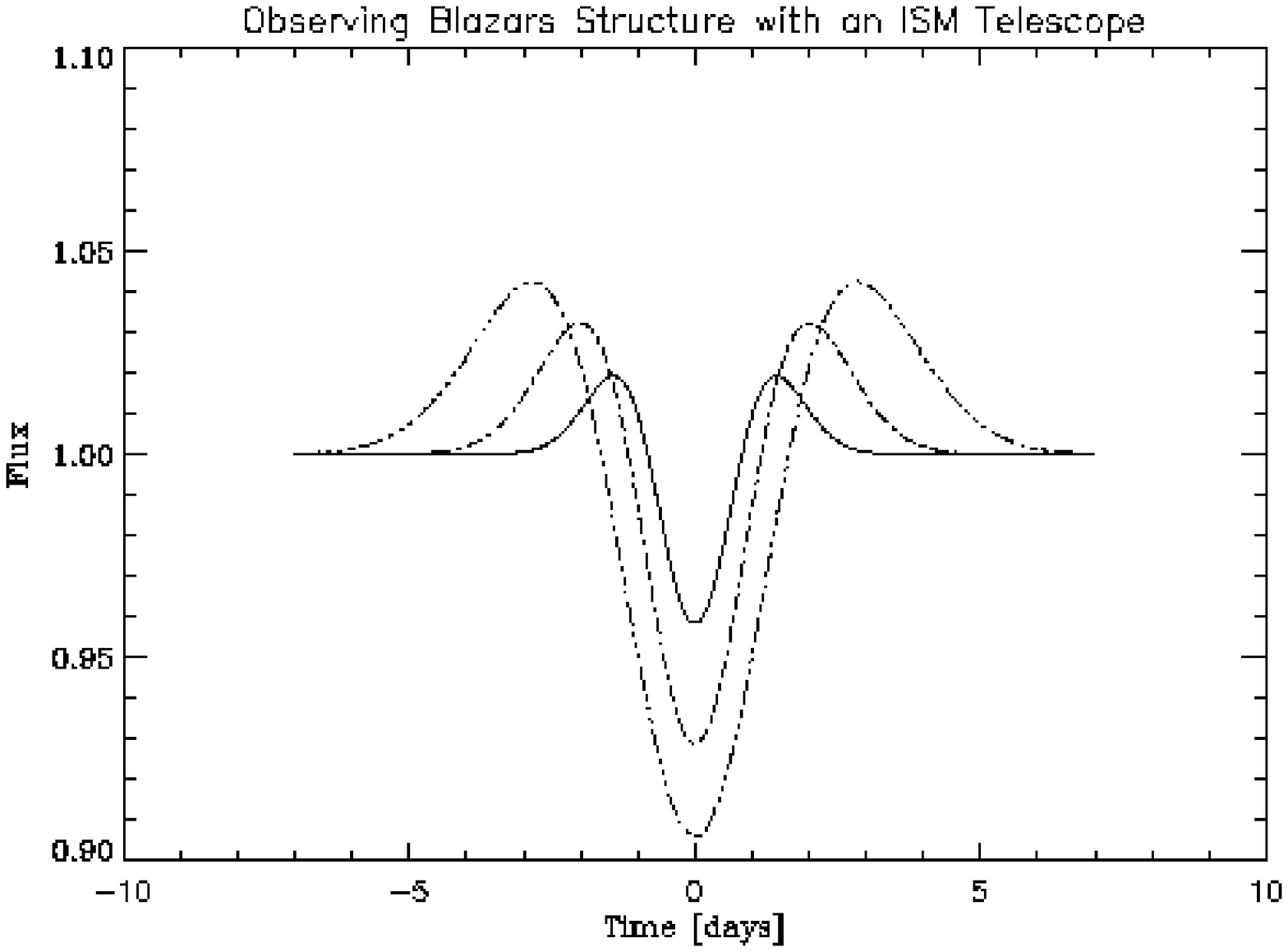,height=5.2cm}
}
\end{minipage}
~~~
\begin{minipage}[t]{7cm}{
\psfig{file=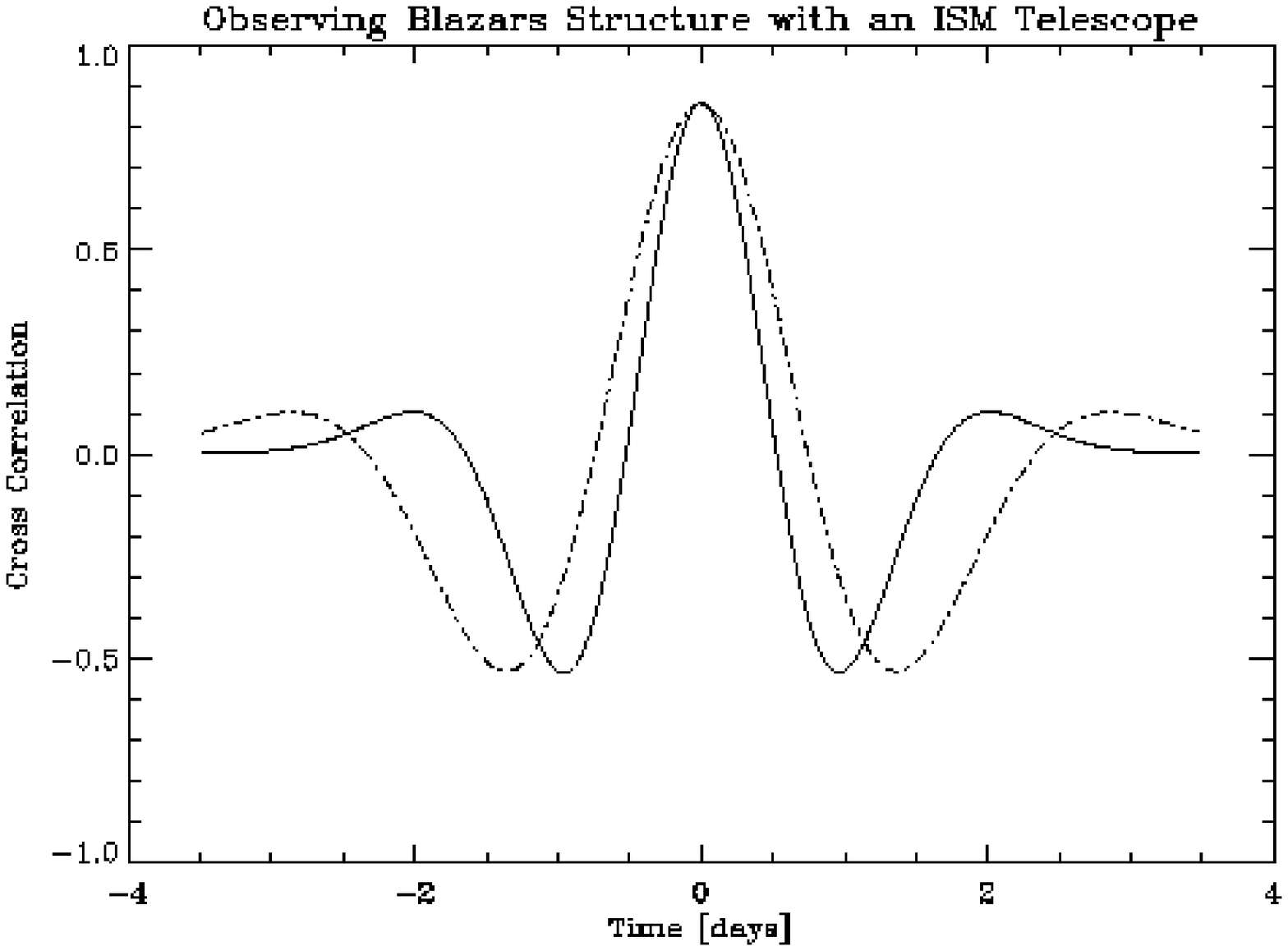,height=5.2cm}
}
\end{minipage}
\caption{\textit{Left:} normalized flux density model light curves
 of a very rapid
ESE at different frequencies, respectively: 11\,cm (dotted line), 6\,cm
(dashed-dotted line) and 3\,cm  (solid line). \textit{Right:} cross
correlation functions for the model light curves for a very rapid ESE
between 11 and 6\,cm (dotted line) and between 6 and 3\,cm (solid line)} 
\label{figlabel3}
\end{figure}

\section{Description of the model and Results}

Galactic density fluctuations along the
line of sight to a compact background extragalactic source are
responsible for Extreme Scattering Events. Such events happen usually
on time scales of weeks to months.
Characteristic of these events is a well defined frequency dependence: 
variability indices, time scales, amplitudes of variations and
amplitudes of the two surrounding maxima decrease with increasing
frequency. Some of these features are similar to normal scattering
processes and the shorter the time scale the more likely is the
confusion of small amplitude ESEs with refractive scintillation.
An important difference between an ESE and 
standard scattering is the focusing, which causes a reversal of the
time lag: observations (Waltman et al. 1991, Fiedler et al. 1994) 
showed that in an ESE the variations (first maximum) start at lower
frequencies. During the minimum, variations at different 
frequencies are simultaneous, then the variations (second maximum)
start at higher frequencies (as we have also observed in our
source). The two bracketing maxima are due to focusing at the edges
of the cloud. A time reversal phenomenon could also appear in the weak
scattering by random variations in the frequency dependence
pattern of RISS (Rickett priv. communication). In the case of an ESE,
we suppose that such a reversal of time lags
is due to changes in the optical depth of the plasma lens.\\ 
Clegg, Fey \& Lazio (1998) developed a plasma lens model describing
ESEs. We extended their model
to small lenses generating very rapid Extreme Scattering Events. 
We applied such a model to the last part of our light curve (box in
fig.\ref{figlabel0}), where
strong variations of the time lag are seen. Fig.\ref{figlabel3} shows
model light curves and cross correlations that we can immediately
compare to our data. In fig.\ref{figlabel1} we plotted the best
fitting of
these models versus the data at 11 and 6\,cm. At 2.8\,cm (not shown
here) the agreement is very similar.\\
A different explanation of time reversal is given by  Qian et
al. (2000), who explain the time reversal in a rapid
outburst of the BL~Lac object 0235+164 by relativistic aberration: a
thin-shock with a large 
Lorentz factor ($\Gamma\geq25$) moves along a curved magnetic field in
the jet.

\section{Discussion}

Assuming a lens speed of 30\,km~s$^{-1}$, the size of the lens would
be in our case 0.035\,A.U.
Considering the lens to be at 0.15\,kpc (distance of Galactic Loop
III, direction close to the line of sight to 0954+658) we can evaluate the
density of the plasma lens: $N_0\approx 1.875\cdot10^{-5}$pc cm$^{-3}$ 
and $n_e\approx 110.95\,\mbox{cm}^{-3}$. Here $N_0$ is the column
density and  $n_e$ the electron density of the cloud.\\
Discussing the ESE in 1981, Clegg, Fey \& Lazio found (at 2.25\,GHz
and assuming the same distance) a lens size of 0.38\,A.U.;
lenses causing very rapid scattering events would require 10
times smaller clouds. For the standard Kolmogorov size
distributions in the ISM, such structures are still physically reasonable,
but they put new constraints on the clumpiness of the
interstellar medium.

\section{Conclusions}

Through a cross correlation analysis, we noted two effects in
the IDV light curve of the source 0954+658.
We suggest that a very rapid ESE (with a time scale of 2 days) occurred
in this source. The comparison of a model for ESEs and our data shows
good agreement, if 10 times smaller clouds than generally accepted 
in the medium are assumed.\\
Previous studies of the IDV in this BL~Lac point towards
an intrinsic explanation for the rapid variations (see Wagner et
al. 1993 for a radio-optical correlation evidence of its IDV), 
even if RISS is present 
due to the compactness of the source. Moreover, the polarized flux
density and polarization angle behaviour of 0954+658 are still hard to 
explain by a simple RISS model. 
In any case, we have to take into account that a clumpy and very turbulent
medium lies along the line of sight to this object. Thus a mixture of
both intrinsic and extrinsic effects is a possible interpretation for
such a complex behaviour.\\
Such a rapid ESE was never observed before. Our analysis can be repeated
on previous data to discriminate between rapid ESE and other sources
of variations.
The above interpretation puts some constraints on the
size (and the density) of the interstellar clouds: the time scale
implies smaller clumpy structures in the ISM than previously thought.

\section*{References}

\reference Clegg, A. W., Fey, A. L. \& Lazio, T. J. W. 1998, ApJ, 496, 253
\reference Fiedler, R. L., Dennison, B., Johnston, K. J. \& Hewish,
A. 1987, Nature, 326, 675
\reference Fiedler, R. L., Dennison, B., Johnston, K. J., Waltman,
E. B., \& Simon, R. S. 1994, ApJ, 430, 581
\reference Heeschen, D. S., Krichbaum, T. P., Schalinski, C. J. \&
Witzel, A. 1987, AJ, 94, 1493
\reference Padovani, P. \& Giommi, P. 1995, MNRAS, 277, 1477
\reference Qian, S. J., Kraus, A., Witzel, A., Krichbaum, T. P. \& Zensus,
J. A. 2000, ApJ, 357, 84
\reference Quirrenbach, A., Witzel, A., Krichbaum T. P., Hummel,
C. A., Wegner, R., Schalinski, C. J., Ott, M., Alberdi, A. \& Rioja,
M. 1992, A\&A, 258, 279
\reference Stickel, M., Fried, J. W. \& Kuhr, H. 1993, A\&AS, 98, 393
\reference Waltman, E. B., Fiedler, R. L., Johnston, K. J., Spencer, 
J. H., Florkowski, D. R., Josties, F. J., McCarthy, D. D. \& Matsakis,
D. N. 1991, ApJS, 77, 379
\reference Wagner, S. J., Witzel, A., Krichbaum, T. P., Wegner, R.,
Quirrenbach, A., Anton, K., Erkens, U., Khanna, R. \& Zensus,
A. 1993, A\&A, 271, 344
\reference Wagner, S. J. \& Witzel, A. 1995, ARA\&A, 33, 163
\end{document}